\documentclass[epj-spec]{svjour}
\usepackage{graphics}
\begin{document}
\title{Film transitions of receding contact lines}

\author{J. Ziegler
\and J.H. Snoeijer
\and J. Eggers
}
\institute{School of Mathematics, 
University of Bristol, University Walk, 
Bristol BS8 1TW, United Kingdom}
\abstract{
When a solid plate is withdrawn from a liquid bath, a receding contact line
is formed where solid, liquid, and gas meet. Above a critical speed $U_{cr}$, 
a stationary contact line can no longer exist and the solid will eventually 
be covered completely by a liquid film. 
Here we show that the bifurcation diagram of this coating transition 
changes qualitatively, from discontinuous to continuous, when 
decreasing the inclination angle $\theta_p$ of the plate. 
We show that this effect is governed by the presence of capillary 
waves, illustrating that the large scale flow strongly effects the 
maximum speed of dewetting.} 

\maketitle

The distinction between a dry solid and a solid covered by a liquid
film is central for all painting and coating processes. In a system
driven either by external motion of the solid or by gravity, this 
distinction is determined not only by the equilibrium properties of 
contact lines, but crucially by non-equilibrium solutions of moving 
contact lines. As a prototypical problem, consider a solid plate 
partially submerged in a liquid, which does not wet the solid. At 
rest, the fluid will form a static meniscus terminating at an angle
corresponding to the equilibrium contact angle $\theta_e$. 
A classical calculation due to Laplace \cite{L05,LL84a} shows that the contact 
line rises (or falls) to a position of $z$ relative to the equilibrium level
of the bath:
\begin{equation}
z = \pm \ell_c\sqrt{2\left[1-\cos \left( \theta_p - \theta_e \right) \right]},
\label{laplace} 
\end{equation}
where $\theta_p$ is the plate inclination, 
$\ell_c = \sqrt{\gamma/(\rho g)}$ is the capillary length, 
and $\gamma, \rho$ the surface tension and the density of the 
liquid, respectively. The $\pm$ sign depends on whether 
$\theta_e$ is smaller (+) or larger (-) than the plate inclination.

\begin{figure}
\begin{center}
\resizebox{0.75\columnwidth}{!}{\includegraphics{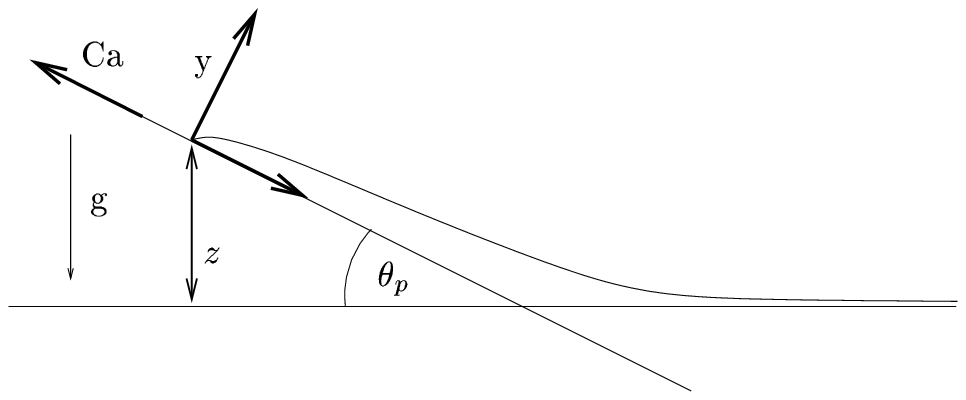}} 
\caption{A solid plate is withdrawn at constant speed from a 
bath of partially wetting liquid. The position of the meniscus 
is measured by $z$, the elevation over the bath.
                 }
\label{fig:1}
\end{center}
\end{figure}

If the plate is withdrawn from the liquid at a constant speed $U$, viscous 
forces will draw up the liquid, and the contact line position 
rises to a new, non-equilibrium value (cf. Fig.~\ref{fig:1}). 
This value results from the competition 
between viscous and capillary forces, so a proper dimensionless measure
of $U$ is the capillary number $Ca=\eta U/\gamma$, where $\eta$ is the
viscosity of the fluid. Above a critical value $Ca_{cr}$ of the capillary 
number (typically 
$Ca_{cr}\mbox{\ \raisebox{-.9ex}{$\stackrel{\textstyle <}{\sim}$}\ } 0.01$),
a stationary contact line position is no longer sustainable, and in the 
long-time limit the plate is covered by a liquid film. According 
to classical phenomenological ideas \cite{DL64}, this transition occurs
when the apparent contact angle, as observed macroscopically, goes
to zero. From (\ref{laplace}) this transition would thus occur as 
the contact line position reaches its maximal height 
$z_0=\ell_c \sqrt{2(1-\cos\theta_p)}$, which is
$\approx \ell_c\theta_p$ for small $\theta_p$.

This idea has recently been confirmed using asymptotic 
matching between the static solution (\ref{laplace}) and the local
fluid motion near the contact line \cite{E04b,E05}. 
The local motion is characterised by a microscopic length $\lambda$, 
which regularises the viscous stress singularity predicted by the 
Navier-Stokes 
equation. For $\lambda = 0$, no contact line motion would be possible
\cite{HS71}. The physical origin of $\lambda$ may depend on the 
physical system at hand, but in this paper we are going to assume 
that the cut-off is due to fluid slip \cite{H83,QWS06}. Typical 
values for $\lambda$ are a few nanometres \cite{CCSC05}. 

However, in principle there is a smaller speed $Ca^*$ 
which is sufficient to initiate the entrainment of a liquid film. 
Snoeijer et al. \cite{SDFA06,SADF07} have investigated the intermediate 
situation of a liquid film that covers a partially dry substrate, 
and thus ends at a receding contact line. 
In the limit that the film is very long, its properties must
become independent of the liquid meniscus, and the tip moves at a
speed $Ca^*$ that only depends on the static contact angle $\theta_e$
and the slip length $\lambda$. Thus, if the plate speed is greater than 
$Ca^*$, a film can be entrained, given appropriate initial conditions. 

In this paper we investigate the bifurcation diagram of the 
entrainment transition as function of the plate angle $\theta_p$. 
This comes about since $Ca_{cr}$ and $Ca^*$ have a different dependence on 
the inclination of the plate. 
Above a critical value of $\theta_p^{cr}$, $Ca_{cr} > Ca^*$, 
and the transition towards the coating layer is discontinuous. 
Namely, once the speed is raised (slightly) above $Ca_{cr}$,
the contact line moves up at a {\it finite} speed $Ca_{cr} - Ca^*$. 
At $\theta_p^{cr}$, however, $Ca_{cr}$ drops below $Ca^*$ and the 
film develops as soon as the plate speed is greater than $Ca^*$. 
In this case the speed at which the contact line moves up 
becomes arbitrarily small at threshold. 
Another striking feature is that for small inclinations the maximum 
height of the meniscus no longer obeys the 'zero contact angle' 
argument predicting a finite $z_0$ at the transition. 
We show that the critical height becomes much larger than 
predicted by (\ref{laplace}) 
and even diverges below a critical plate inclination. 

\noindent
{\bf Results.}
We numerically solved for the shapes of stationary menisci, characterised 
by the liquid thickness $h(x)$, using the lubrication approximation 
\cite{ODB97,H01}. This long wavelength expansion remains quantitatively 
accurate when the interface slope remains small, $h' \ll 1$, so we 
consider small plate inclinations only. As the equilibrium contact 
angle enters as a boundary condition $h'=\theta_e$ at the contact line, 
it is convenient to use the rescaled height $\bar{h}=h/\theta_e$, such 
that $\bar{h}'=1$. If we further introduce 
$\bar{\theta}=\theta_p/\theta_e$, $\bar{\lambda}=\lambda/\theta_e$ and 
$\delta = 3 \mathit{Ca}/\theta_e^3$, and dropping overbars, the equation 
for the meniscus becomes \cite{E04b}:

\begin{eqnarray}\label{scaled}
h^{\prime\prime\prime} - h^{\prime} + \theta= \frac{\delta}{h^2+3\lambda^2}.
\end{eqnarray}
All lengths are expressed in the capillary length $\ell_c$, 
typically a millimetre, while we take the slip-length 
$\lambda=3.3\times 10^{-3}$ throughout the paper.
\footnote{This convenient slip law slightly differs from the 
usual Navier slip condition, as the latter yields a diverging 
$h'''$ at the contact line. This choice has been shown to have no 
effect on the macroscopic physics \cite{E04b}.} The boundary conditions 
for this third order equation are thus $h=0$ and $h'=1$ at the contact 
line, and the third condition comes from matching to the bath, 
$h\simeq x \theta$ as $x \rightarrow \infty$. Solutions are 
found numerically using a shooting procedure.

\begin{figure}
\begin{center}
\resizebox{1.0\columnwidth}{!}{\includegraphics{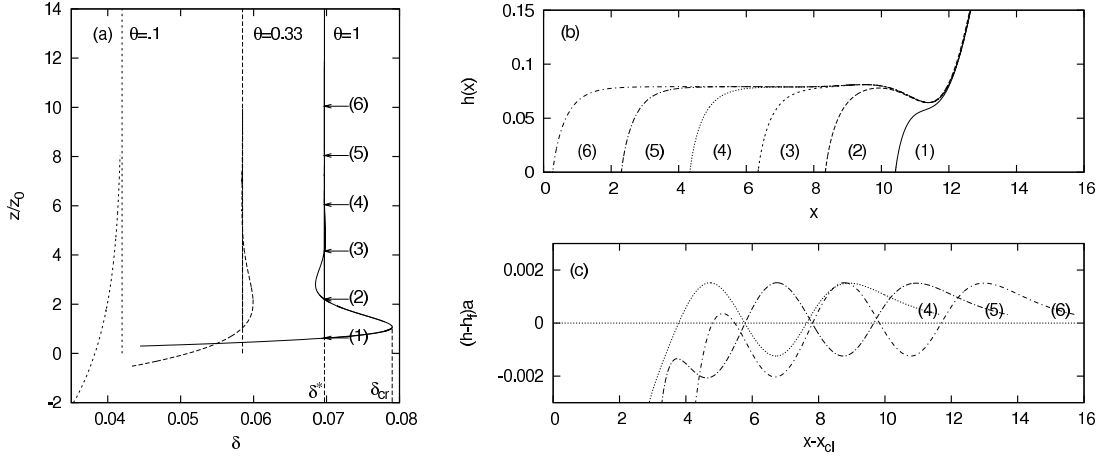}} 
\caption{(a) Bifurcation diagrams: meniscus rise $z$ vs. plate speed 
$\delta$ for various 
plate inclinations $\theta$. At a critical $\theta^{cr}$ the difference 
between $\delta_{cr}-\delta^*$ vanishes and the transition becomes continuous.
(b) Meniscus profiles $h(x)$ for $\theta=1.0$ at locations separated by 
$\ell_{\rm osc}/2$.
(c) To visualise capillary waves on the film, which are exponentially 
damped, we multiplied $(h-h_f)$ by a factor $a=10^{-0.5 (x-x_{cl})}$, 
where $x_{cl}$ is the position of the contact line. The solutions (4) 
and (6) are exactly in phase, whereas (5) is shifted by half a wavelength.} 
\label{fig:dropripple}
\end{center}
\end{figure}

The results are conveniently represented in terms of the meniscus 
elevation $z$, see Fig.~\ref{fig:dropripple} (a). For moderate plate 
inclination ($\theta = 1$), $z$ increases with the (rescaled) plate 
velocity $\delta$, until a saddle-node bifurcation occurs at 
$\delta_{cr}$. The elevation curve then undergoes an infinite series of 
exponentially damped oscillations around another critical speed $\delta^*$, 
each corresponding to a saddle-node bifurcation. The oscillations 
are seen to have a well-defined wavelength $\ell_{osc}$. 
Profiles corresponding to
the marked arrows are shown in panel (b), revealing increasingly 
long films of almost uniform thickness $h_f=\sqrt{\delta^*/\theta}$ 
as one moves up the elevation curve. The natural speed of a 
film, ending in a contact line, is $\delta^*$. As the plate 
velocity is raised above $\delta_{cr}$, a front will start to move
up the plate at a {\it finite} speed $\delta_{cr} - \delta^*$.
However, another scenario observed experimentally \cite{SDFA06} 
is a jump to the film solution as soon as $\delta^*$ is reached. 
The reason for this discontinuous behaviour is not understood at present.

As the plate inclination decreases, the oscillations on the elevation
curve become smaller (cf. Fig.~\ref{fig:dropripple} (a)), and have
vanished entirely for $\theta=0.1$, making the transition toward a 
film continuous. We now show that the oscillations in the bifurcation 
diagram are directly related to small ripples on the actual profiles 
seen in Fig.~\ref{fig:dropripple} (b), and which have {\it the same 
wavelength} $\ell_{wave}$, as seen in the expanded scale, panel (c).
As one moves up the elevation curve by $\ell_{osc}/2$, one 
additional half-wave fits onto the profile, causing successive profiles
in panel (b) to line up exactly, confirming that 
$\ell_{\rm osc}=\ell_{\rm wave}$. One observes that solutions for 
which $\delta$ is at a local maximum (minimum), the oscillation has 
positive (negative) amplitude close to the contact line. This is 
consistent with the fact that dissipation is stronger when the 
thickness is thinner. 

\begin{figure}
\begin{center}
\resizebox{0.9\columnwidth}{!}{\includegraphics{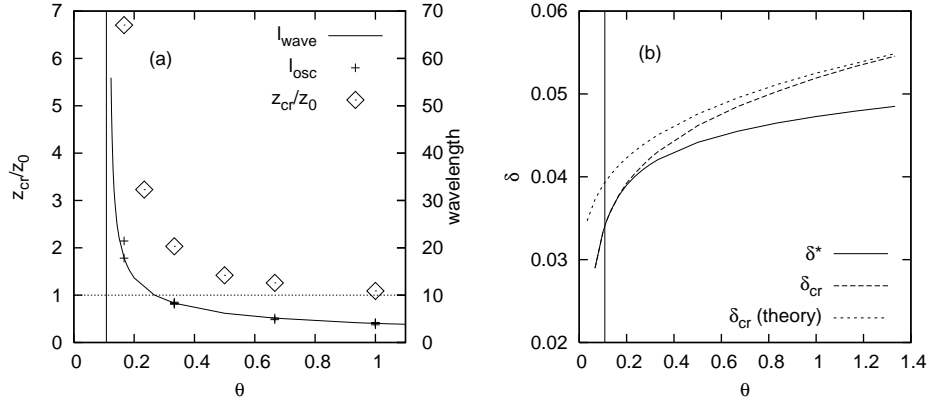}} 
\caption{(a) Right axis: the wavelength $\ell_{osc}$ (+) compared 
to $\ell_{wave}$ (solid line), as computed from (\ref{wavenumber}).
Left axis: comparison of the elevation at the wetting transition, 
$z_{cr}$, to the prediction of zero apparent contact angle, $z_0$
($\diamond$).  
The ratio $z_{cr}/z_0$ diverges as $\theta_{cr}$ is approached.
(b) Dependence of $\delta_{cr}$ (dashed line) 
and $\delta^*$ (solid line) on the plate inclination $\theta$. 
The short-dashed line is the theoretical prediction from asymptotic 
matching \cite{E04b,E05}, which breaks down for small inclinations. 
\label{bifurcation} }
\end{center}
\end{figure}

The above observations imply that the oscillations in the bifurcation
diagram can be understood completely in terms of (small) oscillations
around a liquid film, obtained by linearising (\ref{scaled}) around $h_f$. 
The wavelength $\ell_{wave}$ is easily found to be 
\begin{equation}\label{wavenumber}
\ell_{wave} = \left\{ \begin{array}{rl}
  4\pi / \left(r^{-1/3} - r^{1/3}\right) &  
        \quad \mbox{if $\quad \alpha \geq 1$} \\
  \infty &  \quad \mbox{if $\quad \alpha \leq 1$} 
       \end{array} \right. ,
\end{equation}
where $\alpha = \frac{27 \theta^3}{\delta^*}$ and 
$r=\sqrt{\alpha}-\sqrt{\alpha -1}$. Perfect agreement between 
the analytical result (\ref{wavenumber}) and $\ell_{osc}$ as
measured from the bifurcation diagrams is seen in 
Fig.~\ref{bifurcation} (a) for various plate inclinations $\theta$. 
At a critical value $\theta^{cr} = (\delta^*/27)^{1/3} = 0.108$
the wavelength diverges and oscillations disappear altogether. 
The numerical value will weakly depend on the slip length $\lambda$. 
At the same inclination, at which the film transition
becomes continuous, the elevation $z_{cr}$ at the entrainment 
transition diverges. To further 
test our analytical criterion for this change of behaviour, we
plotted the speeds $\delta_{cr}$ and $\delta^*$ as function of $\theta$
(cf. Fig.~\ref{bifurcation} (b)). 
Indeed, at $\theta = \theta_{cr}$, the two speeds become identical 
and the ``bump'' seen in Fig.~\ref{fig:dropripple} (a) goes away. 

\noindent
{\bf Outlook.}
We have shown that the transition towards liquid deposition can 
be continuous or discontinuous, depending on the inclination of 
the plate. This provides a striking demonstration of the fact that 
maximum speed of dewetting is not an intrinsic property of the 
contact line, as has been claimed in an earlier paper \cite{G86},
but subtly depends on the large scale geometry of the problem. 
A very similar analysis applies to the shape of two-dimensional drops or
ridges sliding down an inclined plane \cite{ZSE07}. Their shapes 
had previously been found numerically, and used as a model problem
to study the stability of contact lines \cite{TK03,K03}. At high
speeds (large volumes) sliding drops acquire a long tail, which 
correspond to the film solutions seen in Fig.~\ref{fig:dropripple} (b).
This contrasts the behaviour of real three-dimensional drops
\cite{PFL01,SLLSE07}, which eventually eject a small rivulet out of 
their rear.

\end{document}